\documentstyle[sprocl]{article}

\begin{document}

\title{EFFECTIVE ACTION FOR DIRAC SPINORS IN THE UNIFORM ELECTROMAGNETIC 
BACKGROUND FIELDS}

\author {ROBERTO SOLDATI}

\address{Dipartimento di Fisica "A. Righi",
                Universit\`a di Bologna,}

\author {LORENZO SORBO}

\address{S.I.S.S.A.-I.S.A.S., Trieste}

\maketitle\abstracts{Some new expressions are found, concerning the effective 
action  as a regularized path-integral for Dirac's spinors in {\it 3+1} 
dimensions, in the presence of general uniform (i.e., constant and 
homogeneous) electric and magnetic fields. The rate of $e^+e^-$ pairs 
production is computed and briefly discussed.}

Path-integrals over Grassmann-valued Fermi
fields are formally given by means of the well known 
Berezin's recipe~\cite{1}.
Gaussian path-integrals of the above kind involving
some self-adjoint
Dirac-like operator  are {\it na\"\i vely} equivalent to 
functional determinants;
for instance
\begin{equation}
\int [d\bar\psi][d\psi]\exp\left\{-\int \bar\psi D\psi dx\right\}=
\prod_n \lambda_n = {\tt det}D,
\label {eq:1}
\end{equation}

where $D=D^\dagger$ and 
$D\psi_n =\lambda_n\psi_n,\quad \lambda_n\not= 0,\
\{\psi_n\}$ being the complete orthonormal set of eigenfunctions of some
self-adjoint invertible operator $D$ with purely discrete non-degenerate 
spectrum.

Functional determinants can be defined in a rigorous mathematical way~\cite{2}
for general dimensionless invertible elliptic systems of partial 
differential operators defined on some compact manifold without boundary.
For such\footnote{The system $A$ is supposed to be elliptic of order
$m>0$, invertible and to admit a ray of minimal grow for its resolvent.}
a system of operators $A$:

{\it (i)}\ {\bf complex} powers $A^{-s}$ do exist,
{\it (ii)}\ $A^{-s}$ has a {\bf continuous kernel} $K_{-s}$,
if the real part of $s$ is sufficiently large,
{\it (iii)}\ $K_{-s}(x,y)$ is an {\bf entire} function of $s$,
where $(x,y)$ are local coordinates on a compact manifold 
${\cal M}$ without boundary and $x\not= y$,
{\it (iv)}\ $K_{-s}(x,x)$ extends to a {\bf meromorphic} function,
{\it (v)}\ $K_{-s}(x,x)$ has {\bf no poles} at $s=0,-1,-2,\ldots$ 

Owing to the above results, we can safely define the $\zeta$-function
associated to the system $A$ as
\begin {equation}
\zeta_A(s)\equiv \int_{{\cal M}} {\tt tr} K_{-s} (x,x) d\mu (x),
\label {eq:2}
\end{equation}
where the trace {\it tr} refers to discrete matrix-like indices of the 
system $A$ while
$d\mu$ is the invariant measure on the compact manifold ${\cal M}$.
Thanks to {\it (v)}, the functional determinant of $A$ can be 
rigorously defined~\cite{3} by means of
\begin{equation}
{\tt det}A\equiv \exp\left\{-\frac{d}{ds}\zeta_A(s=0)\right\}.
\label{eq:3}
\end{equation}

Let us now consider quantized Dirac's four components massive 
spinor fields in the
presence of some background classical uniform (i.e., constant and
homogeneous) general electromagnetic field.
The use of the path-integral method forces us to work in the euclidean 
framework: only at the very end of our calculations we shall
operate a Wick rotation to Minkowski space-time\footnote{On {\bf R}$^4$
one can rigorously define the effective action per unit volume.}.
The effective euclidean 
action in the semi-classical approximation is given by
\begin{equation}
{\cal S}^{\rm E}_{\rm Eff}[A_\mu ]=
\int d^4 x\ \frac{1}{4}F_{\mu\nu} F_{\mu\nu}
-\log {\tt det}\left(\frac{D[A_\mu ]+im}{\mu}\right),
\label{eq:4}
\end{equation}
where 
$D [A_\mu ]\equiv \gamma_\mu (i\partial_\mu +eA_\mu )$
is the self-adjoint 
euclidean Dirac operator, $m$ being the fermion mass and $\mu$ some suitable
mass scale.
As the massive Dirac's 
operator is normal we have
\begin{equation}
\left|{\tt {det}}\left(\frac{D+im}{\mu}\right)\right|^2={\tt {det}}
\left(\frac{D^2+m^2}{\mu^2}\right)\equiv {\tt det}H.
\label{eq:5}
\end{equation}

The effective lagrangean is thereby defined to be
\begin{equation}
{\cal L}^{\rm E}_{\rm {Eff}}[A_\mu ]={\cal L}^{\rm E}_{
\rm {Cl}}[A_\mu ]- \frac{1}{2}\mu^4\left.\frac{\partial}{\partial s}
\zeta_H\left(s\right)\right|_{s=0}.
\label{eq:6}
\end{equation}
Since the square $H$ of the massive Dirac's operator 
is  elliptic, invertible and positive,  its determinant 
can be evaluated by means of the $\zeta$-regularization. 
A frame can always 
be chosen such that 
$F_{03}=-F_{30}=E,\ F_{12}=-F_{21}=B,$
all the other components of $F_{\mu\nu}$ vanishing.

After setting $a\equiv |eE/\mu^2|$, $b\equiv 
|eB/\mu^2|$ and $c\equiv (m^2/\mu^2)$ we get
\begin {equation}
\zeta_H(s)=\frac{ab}{4 \pi^2\Gamma (s)}
\int_0^\infty\ d t\ t^{s-1} {\rm e}^{-ct}
\left\{2 \frac{{\rm e}^{-2at}
+{\rm e}^{-2bt}}{(1-{\rm e}^{-2at})(1-{\rm e}^{-2bt})}+1\right\}.
\label {eq:7}
\end{equation}
Explicit formulae  were known~\cite{4} for 
the effective euclidean lagrangean in the cases
$B=0$ (or $E=0$) and  $E=B$.
It is possible to obtain asymptotic expansions~\cite{5} in powers of $|B/E|$
both in the massive and massless cases.

The starting point is the scaling property
$\zeta_H(s)=a^{2-s} g(s;y,z)$,
so that it is sufficient to study the relevant quantity 
\begin{equation}
g(s;y,z)=
\frac{y}{\Gamma (s)}\int_0^\infty\ d t\ t^{s-1} {\rm e}^{-zt}
\left\{2 \frac{{\rm e}^{-2t}+{\rm e}^{-2yt}}{(1-{\rm e}^{-2t})
(1-{\rm e}^{-2yt})}+1\right\},
\label {eq:8}
\end{equation}
as a function of the ratio 
$y\equiv (b/a)=|B/E|$
at a given $z\equiv (c/a)= |m^2/eE|$.
Expansions around $y=1\Leftrightarrow |E|=|B|$ are easily obtained.
The non-trivial result is the expansion around $|B|=y=0$:
\begin{equation}
\begin{array}{rcl}
g(s;y,z) &=& \frac{2^{1-s}}{\Gamma (s)}\sum_{k=0}^N \frac{{\bf B}_{2k}}
{(2k)!}y^{2k} \Gamma (s-1+2k)
\Bigl \{
2\zeta_R \left(s-1+2k;1+\frac{z}{2}\right)+\\
&+&\left(\frac{z}{2}\right)^{1-2k-s}\Bigr \}
+O\left(y^{2N+2}\right),
\label {eq:9}
\end{array}
\end{equation}
in which ${\bf B}_{2k}$ is the $2k$-th Bernoulli number and
$\zeta_R(a;x)$ is the Riemann-Hurwitz
$\zeta$-function~\cite{6}. 
Exploiting the above scaling property the effective 
euclidean lagrangean of QED in the presence of uniform fields can be easily 
calculated as an asymptotic series in $|B/E|$
or $|E/B|$. 


Now we show how these corrections to the $E=0$ case work, paying 
special attention to the rate of production of fermion-antifermion pairs.
Denoting by $\cal E$ and $\cal B$ the minkowskian electric
and magnetic field strengths respectively,
the transition to the Minkowski space-time is performed by 
means of the substitution
\begin{equation}
{\cal L}_{\rm {Eff}}^{\rm M}({\cal E},{\cal B})=
-{\cal L}_{\rm {Eff}}^{\rm E}(E=-{\rm i}{\cal E},
B={\cal B}).
\label {eq:11}
\end{equation}
The exact pairs production rate per 
unit volume $w= 2Im{\cal L}_{\rm {Eff}}^{\rm M}$
at ${\cal B}=0$ is~\cite{7} 
\begin{equation}
w({\cal E},{\cal B}=0)
=\frac{e^2{\cal E}^2}{4\pi^3}\sum_{n=1}^\infty \frac{1}{n^2}\exp \left\{-n
\frac{\pi m^2}{e{\cal E}}\right\},
\label{eq:13}
\end {equation}
which shows 
that the function $w({\cal E})$ is not analytic in ${\cal E}=0$.
Now, with the 
aid of the previous asymptotic expansion we obtain the first corrections 
of the order $O({\cal B} /{\cal E})^2$ to the total rate 
of production of $e^+e^-$ pairs that reads
\begin{equation}
w=\frac {e^2{\cal E}^2}{4\pi^3}\ \sum_{n=1}^\infty\left(\,\frac{1}{n^2}+
\frac{\pi^2}{3}\,\frac{{\cal B}^2}{{\cal E}^2}\,\right)\exp
\left\{-n\,\frac{\pi m^2}{e{\cal E}}\right\}+O\left(
\frac{{\cal B}^4}{{\cal E}^4}\right).
\label {eq:15}
\end{equation}
The above formulae
are not analytic in ${\cal E}=0$, 
owing to the non-perturbative character of this 
phenomenon.
Considering also
the second term of order $O({\cal B}/{\cal E})^4$ to the 
"unperturbed" effective lagrangean and retaining only the first terms in the 
expansion for weak fields, we get the effective lagrangean in the limit
$|e{\cal B} |\ll|e{\cal E} |\ll m^2/2$: namely,
\begin{equation}
\delta^{(4)}{\cal L}_{\rm {Eff}}^{\rm M}({\cal E},{\cal B}) \simeq 
\frac{e^4}{16\pi^2}\
\frac{2}{45 m^4}\left({\cal E}^4+{\cal B}^4+5{\cal E}^2{\cal B}^2\right) +
\frac{e^2}{24\pi^2}{\cal B}^2\log\frac {m^2}{\mu^2},
\label {eq:16}
\end{equation}
which is in perfect agreement with the Euler-Heisenberg effective 
lagrangean~\cite{8} describing the light-light scattering at very low photons 
momenta, after setting the arbitrary scale 
$\mu$ at the main scale $m$ of the problem.
According to a {\it na\"\i ve} interpretation, the phenomenon of particle 
production in external electromagnetic fields is due to the fact that
virtual pairs created in the vacuum are accelerated by the electric field 
and may gain energy enough to reach the threshold (the electron/positron mass) 
and become physical particles.
On the other hand electrically charged particles do not acquire
energy from a magnetic field, so that
it is rather surprising that a (weak) magnetic field 
gives itself a contribution to the rate of pair creation as we have shown to be
\begin{equation}
(\Delta w)_{{\rm magn}}=\frac{e^2{\cal B}^2}{12\pi}\ 
\sum_{n=1}^\infty\exp
\left\{-n\,\frac{\pi m^2}{e{\cal E}}\right\}+O\left(\frac{{\cal B}^4}
{{\cal E}^4}\right).
\label{eq:17}
\end{equation}



Analogous calculations in the non-Abelian case as well as in the case of
chiral fermions would be quite suggestive and useful.


\section*{References}
\begin{thebibliography}{99}
\bibitem{1}\ F. A. Berezin : "{\it The method of second quantization}",
Academic Press, New York (1966).
\bibitem{2}\ R. T. Seeley : Amer. Math. Soc. Proc. Symp. Pure Math. {\bf 10}
            (1967) 288.
\bibitem{3}\ D. B. Ray, I. M. Singer : Adv. Math. {\bf 7} (1971) 145;
J. S. Dowker, R. Critchley : Phys. Rev. D {\bf 13} (1976) 3224;
S. W. Hawking : Comm. Math. Phys. {\bf 55} (1977) 133.
\bibitem{4}\ S. K. Blau, M. Visser, A. Wipf : Int. Jour. Mod. Phys. A {\bf 6}
            (1991) 5409.
\bibitem{5}\ R. Soldati, L. Sorbo : Phys. Lett. B {\bf 426} (1998) 82.
\bibitem{6}\ A. Erd\'elyi, W. Magnus, F. Oberhettinger, F. G. Tricomi (Eds.) :
            "{\it The Bateman manuscript project: higher transcendental 
            functions}", McGraw-Hill, New York (1953-1955).
\bibitem{7}\ C. Itzykson, J. B. Zuber : "{\it Quantum field theory}",
            McGraw-Hill, New York (1980).
\bibitem{8}\ H. Euler, W. Heisenberg : Z. Phys. {\bf 98} (1936) 714.
\end{thebibliography}
\end{document}